\documentclass[MSNbibl,dvips]{arxstspdf}
\usepackage{flushend}
\usepackage{stfloats}
\usepackage{graphicx}

\volume{26}
\issue{1}
\pubyear{2011}
\firstpage{49}
\lastpage{52}
\doi{10.1214/11-STS345C}
\referstodoi{10.1214/10-STS345}

\makeatletter
\newcommand{\APEone}{APE(${\le} 1$)}
\newcommand{\APET}{APE(${\le} T$)}
\newcommand{\days}{\,\mathrm{day}}
\makeatother

\begin{document}
\begin{frontmatter}

\vspace*{12pt}
\title{Discussion of ``Feature Matching in Time Series Modeling'' by Y. Xia and H. Tong}
\runtitle{Discussion}
\pdftitle{Discussion of Feature Matching in Time Series Modeling by Y. Xia and H. Tong}

\begin{aug}
\author{\fnms{Edward L.} \snm{Ionides}\corref{}\ead[label=e1]{ionides@umich.edu}}
\runauthor{E. Ionides}

\affiliation{University of Michigan}

\address{Edward L. Ionides is Associate Professor, Department of Statistics,
University of Michigan, Ann Arbor,
Michigan 48109, USA \printead{e1}.}

\end{aug}



\end{frontmatter}

Xia and Tong have made a novel contribution to the debate on whether
and how to carry out some sort of feature matching in preference to a
statistically efficient alternative such as the maximum likelihood
estimate (MLE). They show that an estimation criterion emphasizing
long-term predictions has some advantages over the MLE on some
misspecified time series models. However, emphasizing long-term
predictions must lead to a down-weighting of higher-frequency
information in the data. In particular, Xia and Tong's catch-all
approach does not typically share the statistical efficiency of MLE
when the model fits the data adequately. Further, it is necessarily the
case (whatever fitting method is used) that some scientific inferences
one might wish to conclude from fitting a misspecified model are
statistically invalid. Scientific interpretation of fitted parameter
values and predictions using a model that is a statistically poor match
to the data therefore requires considerable care. One seeks models that
are simultaneously scientifically relevant and provide an adequate
statistical description of the data, and then statistical efficiency
becomes an important consideration for drawing scientific conclusions
from limited data. Flexible modern inference methods facilitate the
development and statistical analysis of such models. I will discuss
these issues in the context of Xia and Tong's analysis of Nicholson's
blowfly data. Similar considerations arise in their measles example,
and have been investigated by He, Ionides and King (\citeyear{he10}).

\begin{figure*}

\includegraphics{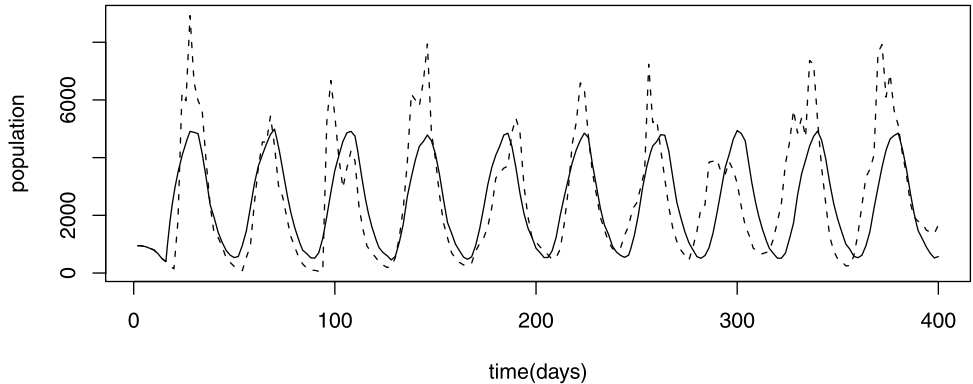}

\caption{Deterministic skeleton for the partially observed stochastic
dynamic model fitted to Nicholson's blowfly data by maximum likelihood
(solid line) and the data (dashed line).}\label{fig:skel}
\vspace*{-6pt}
\end{figure*}

Xia and Tong's {\APEone} estimate is equivalent to the MLE only for a
specific choice of stochastic model. From their equation (3.12), we see
that {\APEone} corresponds to the MLE for additive,\break Gaussian,
constant-variance process noise with no measurement error. For Xia and
Tong's blowfly mo\-del, the log-likelihood at the {\APEone} point
estimate is $-1568.5$ whereas the log-likelihood at the {\APET} point
estimate is $-1569.5$. A chi-squared approximation indicates that a
full likelihood-based analysis for this model should consider the
{\APEone} and {\APET} point estimates to be both statistically
plausible, since the difference of $1.0$ log units is not large
compared to typical values of $1/2$ of a chi-squared random variable
with five degrees of freedom. To check the extent to which either of
these point estimates provides a reasonable statistical explanation of
the data, I compared their goodness of fit with that of a simple
phenomenological model. For oscillating populations, a $\operatorname{log\mbox{-}ARMA}$ model
is an appropriate choice (He, Ionides and King, \citeyear{he10}). I fitted a
stationary $\operatorname{log\mbox{-}ARMA}$ model to the 9th through 200th data points for
which predictions are made by Xia and Tong's model, in order to ensure
that the resulting likelihood provides a fair comparison. A
$\operatorname{log\mbox{-}ARMA}(2,2)$ model gives a maximized log-likelihood of $-1542.3$ based
on estimating six parameters. Xia and Tong's mechanistic model
therefore explains the data considerably more poorly (e.g., judged by
Akaike's information criterion) than this simple black-box model. Is it
possible to preserve the scientific interpretability of Xia and Tong's
model while also providing a statistically satisfactory explanation of
the data? To address this question, I~fitted a dynamic model adapted
from Wood (\citeyear{wood10}) which has a similar structure to the model of Xia
and Tong but differs by formulating the stochasticity in a
scientifically motivated way. This alternative model is described in
full in the Appendix below. I~evaluated the likelihood by sequential
Monte~Carlo and computed the MLE by iterated filtering
(Ionides, Bret{\'{o}} and
King, \citeyear{ionides06pnas}) implemented using the {\texttt{pomp}} package for
R (King et al., \citeyear{king10pomp}). Maximization over the six parameters led to a
log-likelihood of $-1465.4$. Figure \ref{fig:skel} shows that the
skeleton of this alternative model matches the periodicity in the data,
a~measure of fit which Xia and Tong chose to emphasize in their
Figure~8. The likelihood at the MLE also comfortably outperforms the
$\operatorname{log\mbox{-}ARMA}(2,2)$ benchmark so subsequent analysis can consider the model
to be adequately specified, at least to a first approximation. Of
course, the possibility of potential further advances in the model
specification cannot be ruled out. Indeed, a~careful and complete
investigation would be expected to reveal some aspects of this improved
model that are a statistically significant mismatch to the data.

The analysis above may suggest that choice of~mo\-del is more important
than the specific choice of inference methodology.
However, model development is facilitated by statistical methodology
that is appropriate for general classes of models (so that the
scientist is not constrained by the methodology when developing models)
and which is convenient for quantitative comparisons between models.
Xia and Tong's {\APEone} and {\APET} criteria are not appropriate for
nonstationary, partially observed dynamic systems evolving in
continuous time.
These features are typical of ecological and epidemiological systems
(Bj{\o}rnstad and Grenfell, \citeyear{bjornstad01}).
Likelihood is quite generally applicable in theory, though
feature-ma\-tching methodology has previously been advocated to avoid the
practical numerical issues of working with the likelihood for dynamic
models (Wood, \citeyear{wood10}, and references therein).
Recently, calculation and maximization of the likelihood function for
general nonlinear, partially observed dynamic models has become
computationally routine in many ecological dynamic systems (e.g., King et al., \citeyear{king08};
He, Ionides and King, \citeyear{he10}; Laneri et al., \citeyear{laneri10}).

A criterion such as {\APET} may help to emphasize certain low-frequency
(long time scale) features of the data such as\vadjust{\goodbreak} the periodicities in the
blowfly population.
While this may be of scientific interest as a component of a data
analysis, it is not desirable as a complete analysis due to the obverse
property of suppressing high-frequency (short time scale) features.
The efficiency of the MLE corresponds to an optimal balance between
frequencies, in the specific sense of minimizing asymptotic variance of
parameter estimates when the model is correct.
This balance between frequencies is perhaps most clearly seen in the
context of Whittle's approximation to the likelihood, discussed by Xia
and Tong in Section 2.2.
Although the usual decomposition of the likelihood for dynamic models
appears to emphasize one-step prediction, the combination of all
one-step predictions corresponds to an estimator which efficiently
combines the contributions of all frequencies.
I shall argue that high-frequency features may be potentially even more
scientifically important than low-frequency features.

Both the blowfly and measles examples involve analyzing mechanistic
models that aim to explain the long-term dynamics of the system in
terms of models constructed to describe the short-term increments or
temporal derivatives (Brillinger, \citeyear{brillinger08}; Bret{\'o} et al., \citeyear{breto09}).
The {\APET} estimate necessarily has a poorer fit than the one-step
{\APEone} estimate, in a least squares sense, to the short-term
behavior that provides the scientific rationale for the mechanistic model.
Xia and Tong's blowfly example suggests that this property can lead to
a greater scientific interpretability of the {\APEone} parameter
estimates. I consider each parameter in turn:
\begin{enumerate}
\item
In the biological interpretation of Xia and Tong's model, $c$
corresponds to the number of eggs laid per adult blowfly per bi-day
that develop into adults in\vadjust{\goodbreak} the absence of competition for food.
From data on eggs collected in this blowfly experiment (Brillinger et al., \citeyear{brillinger80})
we see that this biological quantity peaked at $c\approx
20$ in the troughs of the population cycles.
This matches closely the estimate $\hat c_1=20.1$ via the {\APEone} criterion.
The {\APET} estimate, $\hat c_T=592$, is an order of magnitude higher
than this biological interpretation permits.

\item
The original biological motivation for Xia and Tong's model had $\alpha
=1$ (Gurney, Blythe and Nisbet, \citeyear{gurney80}) and a value slightly less than 1 has been proposed
when making a discrete-time approximation to a continuous dynamic system.
The {\APEone} estimate $\hat\alpha_1=0.846$ is consistent with this
interpretation, whereas the {\APET} estimate $\hat\alpha_T=0.263$ is so
far below unity that it requires a reinterpretation of the biological
story behind the model.

\item
Biologically, $\alpha N_0$ is the adult population size that maximizes
the total number of successfully-\break developing eggs laid.
Empirically, the adult population size maximizing total egg production
occurred during troughs of adult abundance at successive values of 397,
542, 167, 2236, 2267, 539, 1308, 2363, 3806 and 254 adults for the ten
cycles analyzed.
The {\APEone} estimate $\hat\alpha_1 \hat N_{0,1}=499$ and the
{\APET} estimate $\hat\alpha_T \hat N_{0,T}=344$ are both broadly
consistent with this interpretation.

\item
$2/(1-\nu)$ may be biologically interpreted as the life expectancy of
the blowfly adults.
The estimates $2/(1-\hat\nu_1)=8.33$ and $2/(1-\hat\nu_T)=5.67$ are
both broadly biologically plausible.
Empirically, life expectancy decreased substantially when the adult
population was large (Brillinger et al., \citeyear{brillinger80}; Guttorp, \citeyear{guttorp81}), and so one must
permit some flexibility in the interpretation of the constant life
expectancy assumed by this model.
\end{enumerate}

In conclusion, Xia and Tong's {\APEone} and\break {\APET} fits to the blowfly
data are statistically more-or-less equally valid.
Both are handicapped by the substantial misspecification of the fitted model.
The {\APET} estimate fits the periodicity of the fluctuations better
but at the expense of the biological interpretation of the fitted parameters.
Superior models can simultaneously satisfy each of these considerations.
If the model is adequately specified, likelihood-based analysis
provides a powerful set of tools for investigating the range of
statistically plausible parameter values.
If the model is poorly specified, likelihood provides a powerful
framework for diagnosing the misspecification and a flexible framework
for constructing improved models.

\appendix
\section*{Appendix: An Alternative Blowfly Data Analysis}
\vspace*{3pt}

Let $N(t)$ be the number of adult blowflies at time~$t$.
Suppose that the number of newly emerging adults during the time
interval $[t,t+\Delta]$ is $R_t$, and the number of adults surviving
from time $t$ to $t+\Delta$ is $S_t$, so that $N(t+\Delta)=R_t+S_t$.
Suppose that $R_t$ and $S_t$ are conditionally independent given $N(t)$
and $N(t-\tau)$ with conditional distributions
\begin{eqnarray*}
R_t&\sim& \operatorname{Poisson}[ N(t-\tau) P\, \exp\{-N(t-\tau)/
N_0\} \, \Delta\, e_t],
\\
S_t&\sim&\operatorname{Binomial}[
N(t),\exp\{- \delta\, \Delta\, \varepsilon_t\} ].
\end{eqnarray*}
Here, $e_t$ and $\varepsilon_t$ are independent Gamma-distributed
random effects with mean $1$ having respective variances $\sigma
_p^2\Delta^{-1}$ and $\sigma_d^2\Delta^{-1}$.
When $\Delta= 2 \days$ this mo\-del is similar to the model of Xia and
Tong, with parameters $N_0$ and $\tau$ having matching interpretations
and the remaining parameters translating to $\alpha= 1$, $c \approx
2P$ and $\nu\approx\exp(-2\delta)$.
When $\Delta= 1 \days$ this corresponds exactly to the dynamic model
of Wood (\citeyear{wood10}).
Wood (\citeyear{wood10}) employed a generalized method of simulated moments to
estimate parameters, but I shall instead construct a partially observed
Markov process (POMP) model for which likelihood-based methods are available.

Supposing that $\Delta$ is chosen to divide $\tau$, the above
construction defines a discrete-time Markov process $X(t)=
(N(t),N(t-\Delta),N(t-2\Delta),\ldots,N(t-\tau))$.
The choices $\Delta=1\days$ and $\Delta=2\days$ can then be viewed as
Euler approximations to a continuous-time Markov process that is
defined by taking the~li\-mit $\Delta\to0$ (Bret{\'o} et~al., \citeyear{breto09}).
To complete a~POMP model, one needs to specify initial conditions and a
measurement process.
Write Nicholson's recorded data as $y_1,\dots, y_T$ where $y_k$ gives
the adult blowflies counted at time $t_k=2k \days$, and $T=200$.
For comparison with Xia and Tong, I fixed $\tau=14\days$ and required
that the model should provide a likelihood for $y_9,y_{10},\dots,y_T$.
The initial state $X(t_8)$ can be constructed using $y_1,\dots,y_8$.
With $\Delta=2\days$, I chose to set $N(t_k)=y_k$ for $k\in\{1,\dots,8\}
$ rather than treating the initial conditions as unknown parameters.
For general $\Delta$, I specified $X(t_8)$ using a cubic spline
interpolation of $y_1,\dots,y_8$.

My measurement model was $y_k\sim\operatorname{Negbinom}( N(t_k),\break\sigma
_y^{-2} )$, a negative binomial distribution conditional on
$N(t_k)$ with mean $N(t_k)$ and variance $N(t_k)+[\, \sigma_y N(t_k) ]^2$.
Nicholson's adult blowfly counts certain\-ly contained some error due to
an inconsistency between the counts of dead adults and newly emerging
adults that were used to infer the counts of living adults (Brillinger et~al., \citeyear{brillinger80}).
However, the uncertainty in the measurement model necessary to provide
a good statistical fit to the data has a more subtle interpretation.
The fertility of adults varies according to their age and potentially
for other unmodeled biological reasons.
In the scientific motivation of the process model, the process model
for $N(t)$ may be interpreted as describing fertility (in units of
ideal, standardized adults) rather than simply measuring the actual
number of adults.
The measurement error then includes fluctuations in the calibration
between the actual number of adults present and their reproductive potential.

Likelihood-based inference for POMP models using iterated filtering has
been described and discus\-sed elsewhere (Ionides, Bret{\'{o}} and
King, \citeyear{ionides06pnas};~King et al., \citeyear{king08};
Ionides, Bret{\'{o}} and
King, \citeyear{ionides08chapter}; Bret{\'o} et~al., \citeyear{breto09};
Bhadra, \citeyear{bhadra10}; He, Ionides and King, \citeyear{he10}; Laneri et~al., \citeyear{laneri10}).
This methodology involves employing sequential Monte Carlo techniques
for eva\-luation and optimization of the likelihood function.
The dynamic process model enters the computations only through the
generation of sample paths at varying values of the parameters.
Methodology enjoying this property has been called plug-and-play
(Bret{\'o} et~al., \citeyear{breto09}; He, Ionides and King, \citeyear{he10}) since it can be implemented simply by plugging
simulation code for the process model into inference software.
In particular, likelihood-based inference is possible even when the
likelihood function itself can be evaluated only by Monte Carlo methods.

There was some indication that the alternative model fits better for
$\Delta=1\days$ (maximized log-likeli\-hood of $-1465.4$) than for $\Delta
=2\days$ (maximized log-likelihood of $-1471.4$).
I did not investigate the introduction of an exponent $\alpha$ that Xia
and Tong proposed to modify the effect of a large time discretization step.
One of the advantages of the POMP framework is that it applies to
continuous-time process models, or models based on arbitrarily small
time discretizations, which makes such modifications unnecessary (Bret{\'o} et~al., \citeyear{breto09}).
Here, there is little reason to prefer the analysis with $\Delta=2\days
$ to $\Delta=1\days$.
The MLE\vspace*{1pt} for $\Delta=1\days$ was
$\hat P=3.28\days^{-1}$,
$\hat N_0=680$,
$\hat\delta=0.161\days^{-1}$,
$\hat\sigma_p=1.35\days^{1/2}$,
$\hat\sigma_d=0.747\days^{1/2}$
and
$\hat\sigma_y=0.0266$.
All parameters are seen to be consistent with the biological
interpretation of the model. The measurement uncertainty parameter,
$\sigma_y$, is estimated to be small so most of the stochasticity is
assigned to variability in the dynamic process.

\section*{Acknowledgments}

This work was supported by NSF Grant
DMS-08-05533 and by the RAPIDD program at the Science \& Technology
Directorate of the Department of Homeland Security, the Fogarty
International Center, National Institutes of Health. Aaron King
contributed to the development of the alternative blow\-fly data
analysis.

%

\end{document}